\documentclass{PoS}
\usepackage{url}
\def\spose#1{\hbox to 0pt{#1\hss}}
\def\ltapprox{\mathrel{\spose{\lower 3pt\hbox{$\mathchar"218$}}
 \raise 2.0pt\hbox{$\mathchar"13C$}}}
\def\gtapprox{\mathrel{\spose{\lower 3pt\hbox{$\mathchar"218$}}
 \raise 2.0pt\hbox{$\mathchar"13E$}}}
\def\inapprox{\mathrel{\spose{\lower 3pt\hbox{$\mathchar"218$}}
 \raise 2.0pt\hbox{$\mathchar"232$}}}

\title{$K\to\pi\pi$ matrix elements\\ from mixed action lattice QCD}

\ShortTitle{$K\to\pi\pi$ matrix elements from mixed action lattice QCD}

\author{Jack Laiho\\
        Department of Physics and Astronomy, University of Glasgow, Glasgow, Scotland, UK\\
        E-mail: \email{jlaiho@fnal.gov}}
        
\author{Ruth S. Van de Water\\
        Physics Department, Brookhaven National Laboratory, Upton, New York, USA\\
        E-mail: \email{ruthv@bnl.gov}}

\abstract{
We present a new method for determining $K\to\pi\pi$ matrix elements from
 lattice simulations that is less costly than direct
 simulations of $K\to\pi\pi$ at physical kinematics.  It improves, however, upon the traditional ``indirect'' approach of
 constructing the $K\to\pi\pi$ matrix elements using NLO $SU(3)$ $\chi$PT,
which can lead to large higher-order chiral corrections.  Using the explicit example of the $\Delta I =3/2$
$(27,1)$ operator to illustrate the method, we obtain a value for Re($A_2$) that agrees with experiment and has a total uncertainty of $\ltapprox$~20\%. 
Although our simulations use domain-wall valence quarks on the MILC
 asqtad-improved gauge configurations, this method is more general and can be
 applied to calculations with any fermion formulation.}

\FullConference{The XXVIII International Symposium on Lattice Field Theory, Lattice2010\\
		June 14-19, 2010\\
		Villasimius, Italy}

\begin{document}

\section{Motivation}
Lattice calculations of $K\to\pi\pi$ matrix elements are important for understanding the Standard Model and in constraining physics
beyond the Standard Model.    For example, they are needed to explain the origin of the $\Delta I = 1/2$ rule and to compute the long-distance contributions to neutral kaon mixing~\cite{Buras:2008nn}.  Because the lowest-order Standard Model contributions to $\epsilon^\prime/\epsilon$ are from 1-loop electroweak penguin diagrams, $K\to\pi\pi$ decay is sensitive to physics at very high scales.  Many extensions of the Standard Model lead to new particles that enter the loops, and these contributions to  $K\to\pi\pi$ may be sufficiently large that they can be observed once the hadronic uncertainties in the weak matrix elements are small enough.  

\begin{figure}
\begin{center}
\includegraphics[width=0.65\linewidth]{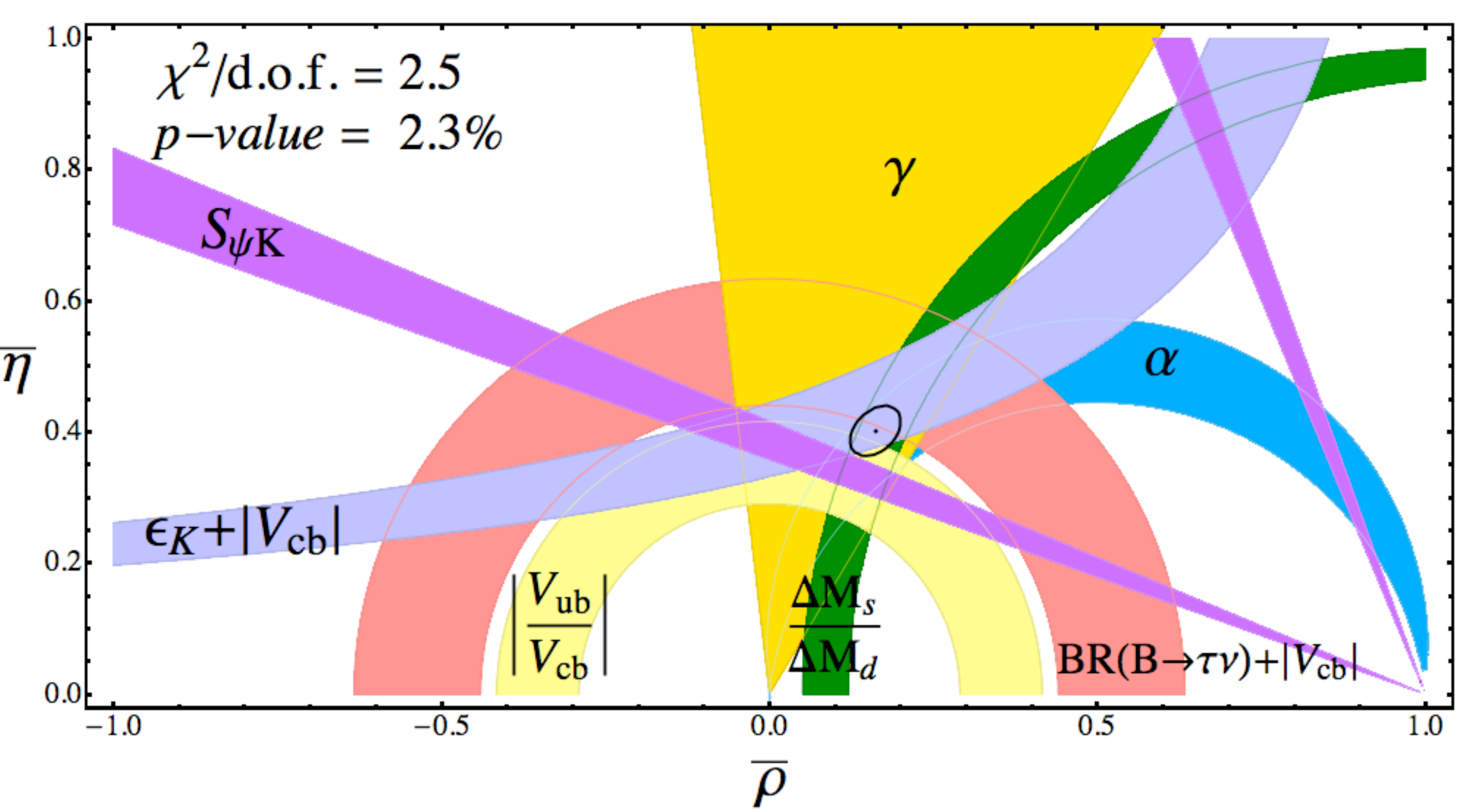}
\vspace{-2mm}
\caption{Global fit of the CKM unitarity triangle~\cite{Laiho:2009eu}.}\label{utfit-full.pdf}
\end{center}
\end{figure}

A standard way of searching for new physics in the flavor sector is by overconstraining the angles and sides of the CKM unitarity triangle~\cite{Antonelli:2009ws}.  This requires precise experimental measurements and equally well-controlled theoretical calculations of hadronic weak matrix elements using lattice QCD.  For many years improved measurements and calculations have simply confirmed the Standard Model CKM framework at the few-percent level, but recent $N_f = 2+1$ flavor lattice calculations of $B_K$ with $\sim$ 4\% precision~\cite{Lubicz:2010nx}  have revealed a 2-3$\sigma$ tension in the CKM unitarity triangle~\cite{Buras:2008nn,Lunghi:2008aa,Laiho:2009eu}.  This tension, which may be due to kaon or $B_d$-meson mixing, is illustrated in Fig.~\ref{utfit-full.pdf}.  Almost all constraints on the CKM unitarity triangle, however, come from the $B$-meson sector.  Thus it is essential to place other constraints on the unitarity triangle from the kaon sector in order to test whether the amount of observed $CP$-violation in the $B$-meson sector is the same as in the kaon sector.  Once lattice QCD calculations of $K\to\pi\pi$ matrix elements are sufficiently precise, they can be combined with the experimental measurement of $\epsilon_K'/\epsilon_K$ to impose an additional constraint on the apex of the CKM unitarity triangle.

\section{New approach to $K\to\pi\pi$ matrix elements}
The Maiani-Testa no-go theorem states that physical matrix elements cannot be extracted from Euclidean correlation functions with multi-hadron states~\cite{Maiani:1990ca}.  Therefore two general approaches have been developed to evade the Maiani-Testa theorem and allow the determination of $K\to\pi\pi$ matrix elements using lattice QCD.  The ``direct'' Lellouch-L\"uscher finite-volume method~\cite{Lellouch:2000pv} is the most straightforward to implement, but it is computationally demanding because it requires a large ($\sim$ 6 fm) box and physical light-quark masses.  The ``indirect'' method constructs $K\to\pi\pi$ matrix elements using the low-energy constants  (LEC's) of $SU(3)$ $\chi$PT obtained from calculating simpler lattice quantities such as $K\to0$  and $K\to\pi$.  Although it was shown that all LEC's through next-to-leading order can be obtained from such ``simple'' lattice quantities~\cite{Laiho:2003uy}, this approach relies on the use of $SU(3)$ $\chi$PT at the kaon mass, where the convergence of the chiral expansion is quite slow.

\begin{figure}
\begin{center}
%
\begin{picture}(147,70) 
\put(-1,0){\includegraphics[width=0.52\linewidth]{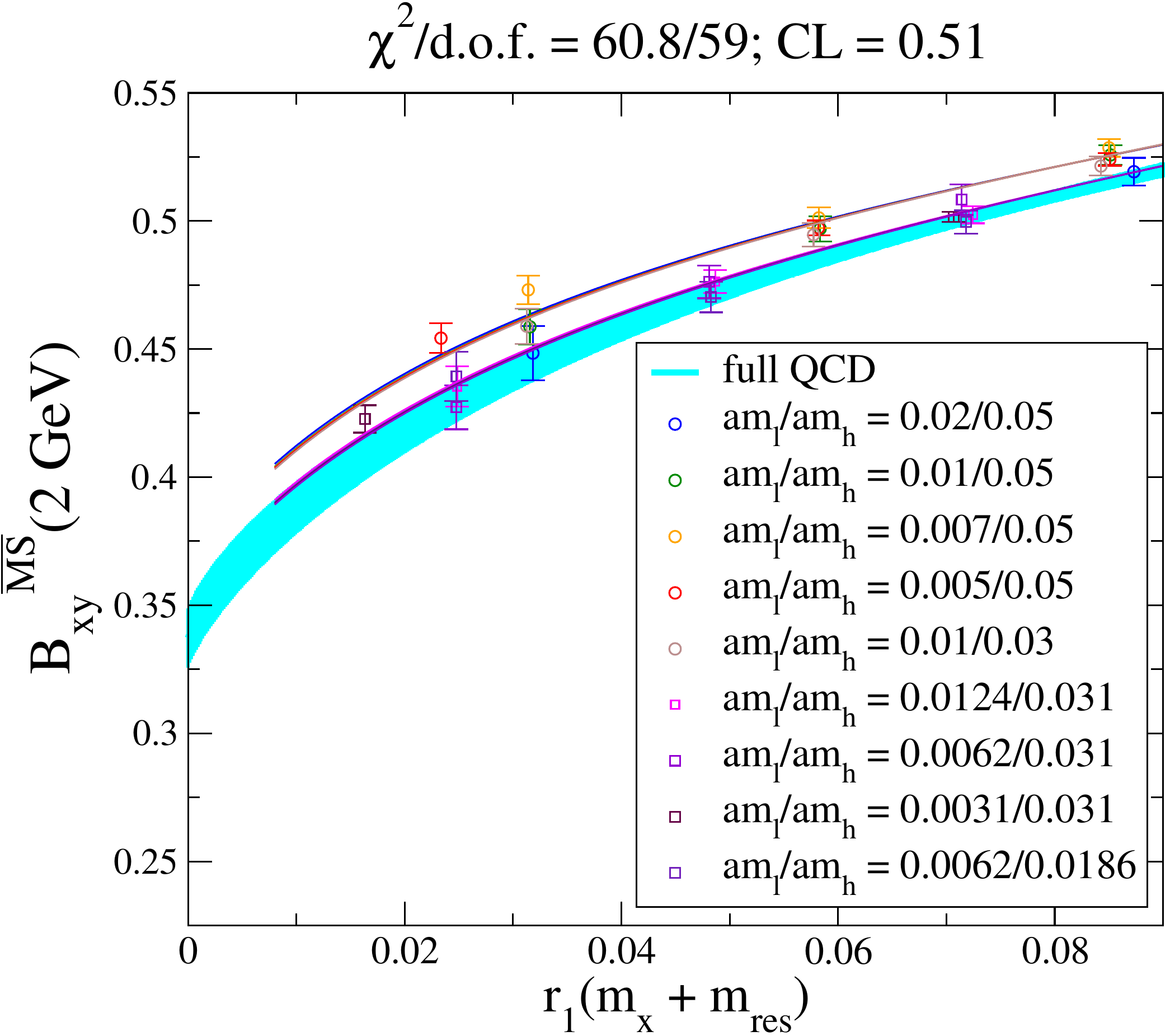}}
\put(80,15){\includegraphics[width=0.45\linewidth]{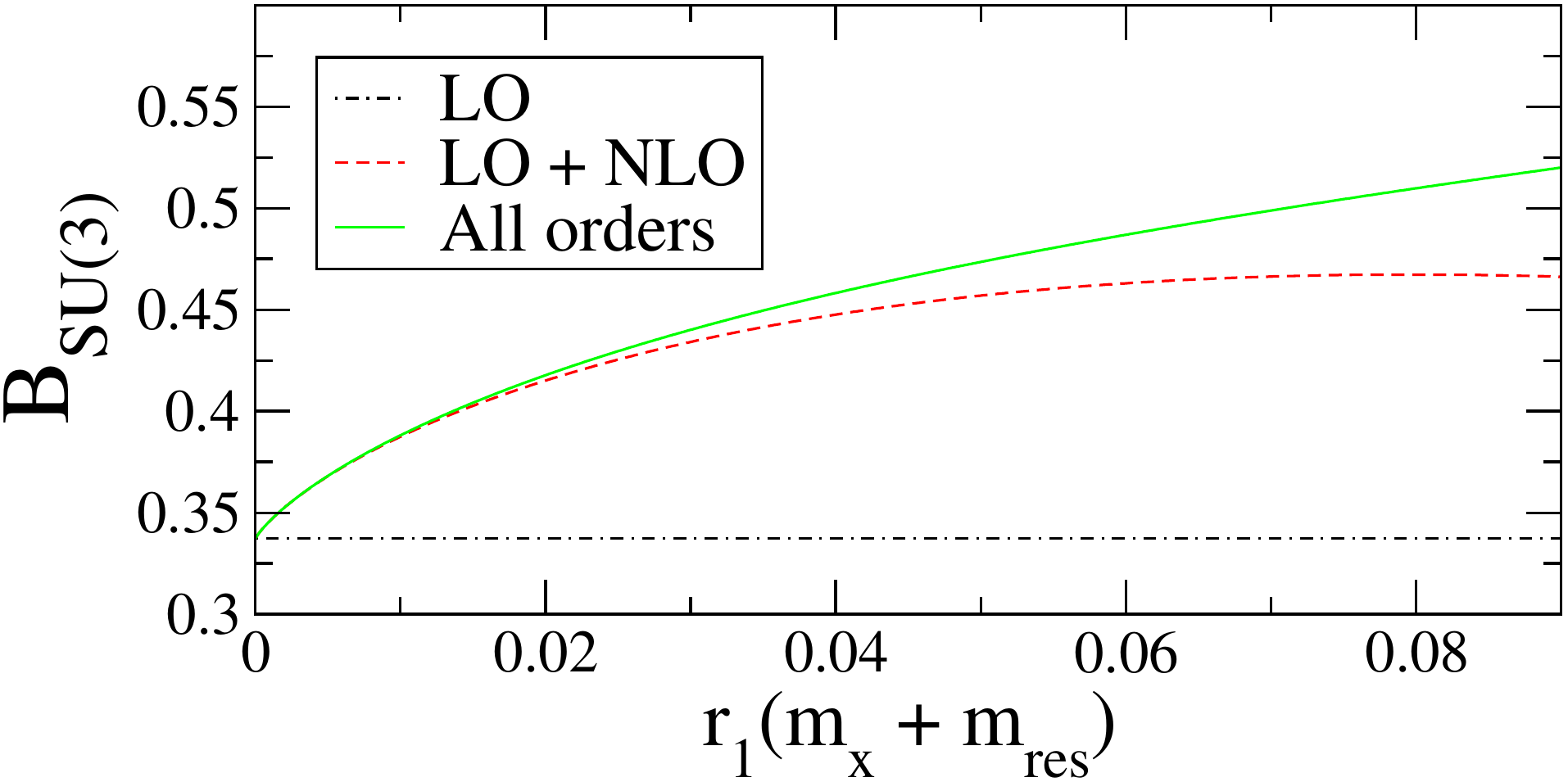}}
\end{picture}
\vspace{-1mm}
\caption{Chiral and continuum extrapolation of $B_K$~\cite{Aubin:2009jh} (left plot) along with convergence of the $SU(3)$ $\chi$PT fit (right plot).   Circles (squares) denote $a \sim 0.12$ fm ($a \sim 0.09$ fm) data.  Only degenerate points are shown, but the fit also includes non-degenerate data.  The cyan band is the degenerate quark mass full QCD curve ($m_x = m_y = m_l = m_h$) in the continuum limit.  The y-intercept of the band gives the LEC $B_0$, the value of $B_K$  in the $SU(3)$ chiral limit.  The right-most point on both plots corresponds to $\sim m_s/2$. $\qquad$}\label{fig:BK_convergence}
\end{center}
\end{figure}

Li and Christ studied the extraction of $K\to\pi\pi$ matrix elements with $N_f = 2+1$ dynamical domain-wall lattice simulations using the ``indirect'' method in Ref.~\cite{Li:2008kc}.  They concluded that large uncertainties in the LO and NLO $SU(3)$ LEC's and the slow convergence of $SU(3)$ $\chi$PT at the scale of the kaon mass lead to large errors that make the extraction of $K\to\pi\pi$ matrix elements using the ``indirect'' method unreliable.  Our procedure therefore addresses these drawbacks of the traditional approach and improves upon it in several ways.  In the combined chiral-continuum extrapolation, we use the physical pseudoscalar meson masses and decay constant.  This leads to better fits as measured by the correlated $\chi^2$/d.o.f.; nontrivial agreement between the NLO mixed-action $\chi$PT prediction for the isovector scalar correlator and lattice simulation data lends support to this approach~\cite{Aubin:2008wk}.  When the fixed-order (NLO) fit is bad, we approximate higher order terms in the chiral expansion by polynomials.  This leads to larger leading-order terms and hence suggests better convergence than was found in Ref.~\cite{Li:2008kc}.  For example, Fig.~\ref{fig:BK_convergence} shows the $SU(3)$ $\chi$PT fit of $B_K$ along with the sizes of the various contributions~\cite{Aubin:2009jh}.  The NLO corrections are only approximately 1/3 of the LO terms even at $m_s/2$.

Despite these findings, however, NLO $\chi$PT corrections can still be 50\% or more for some quantities. Therefore, to achieve the precision needed for $K\to\pi\pi$ we do not rely on the ``indirect'' method alone. Rather, we combine indirect and direct methods in a cost-effective way.  We bypass the Maiani-Testa theorem by simulating with both pions at rest.  We fit the numerical data to NLO mixed-action $\chi$PT plus higher-order analytic terms, extrapolate to the continuum, and interpolate to the point at which $m_K = m_K^{\textrm{phys.}}$ and $m_\pi = m_K^{\textrm{phys.}}/2$.  Thus we avoid relying upon $SU(3)$ $\chi$PT to extrapolate to the physical kaon mass, where we expect higher-order corrections to be significant.  We then correct this unphysical kinematics point using fixed-order $SU(3)$ $\chi$PT. The low energy constants needed for this correction can be obtained from simpler quantities such as $f_K$, $K^0$-$\bar{K^0}$, and $K\to\pi$. Since the kaon is tuned to its physical value, terms involving only kaons are correct to all orders in the $SU(3)$ chiral expansion;  we therefore expect higher-order corrections to be small.

\begin{figure}
\begin{center}
\includegraphics[height=0.27\linewidth]{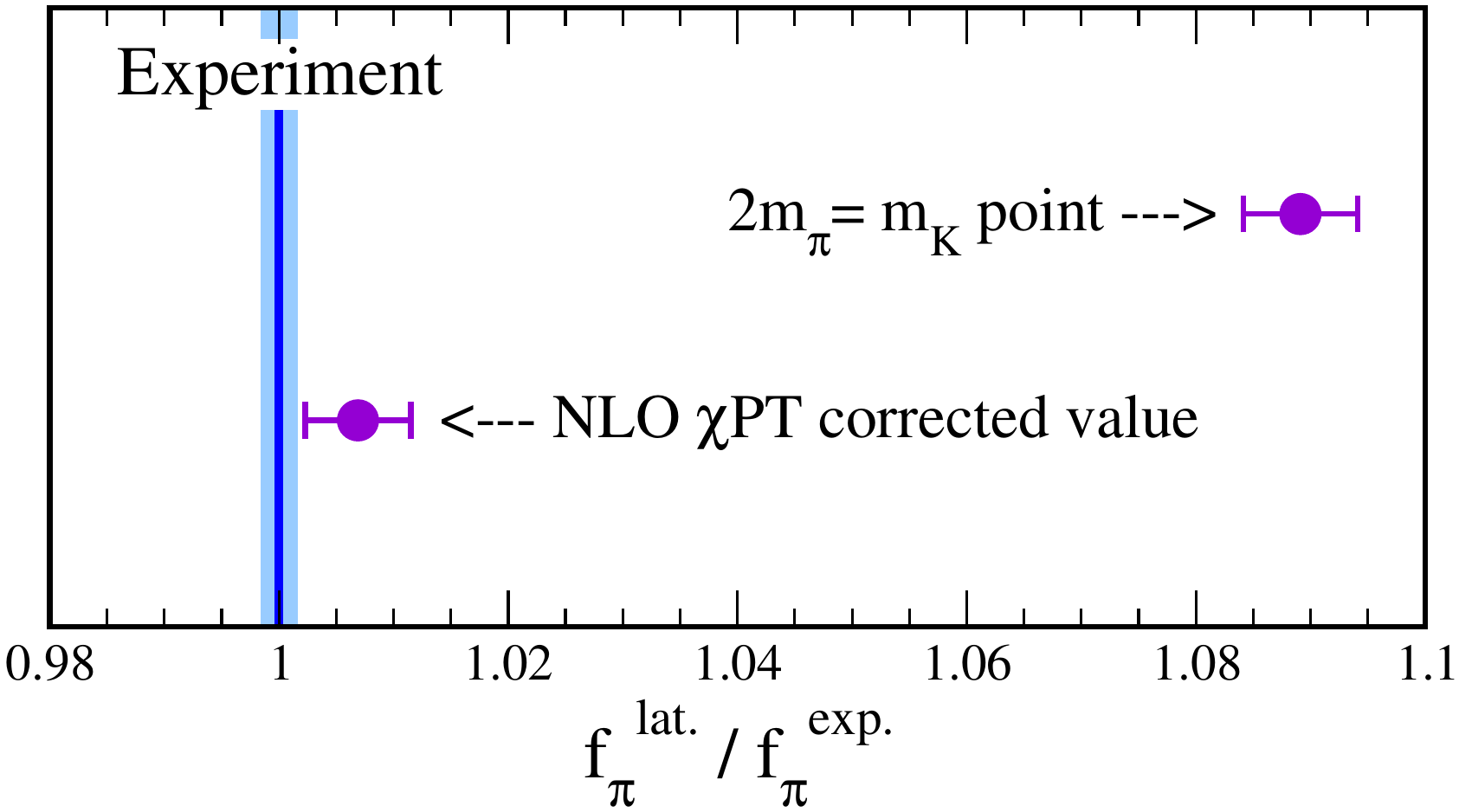} $\quad$
\includegraphics[height=0.27\linewidth]{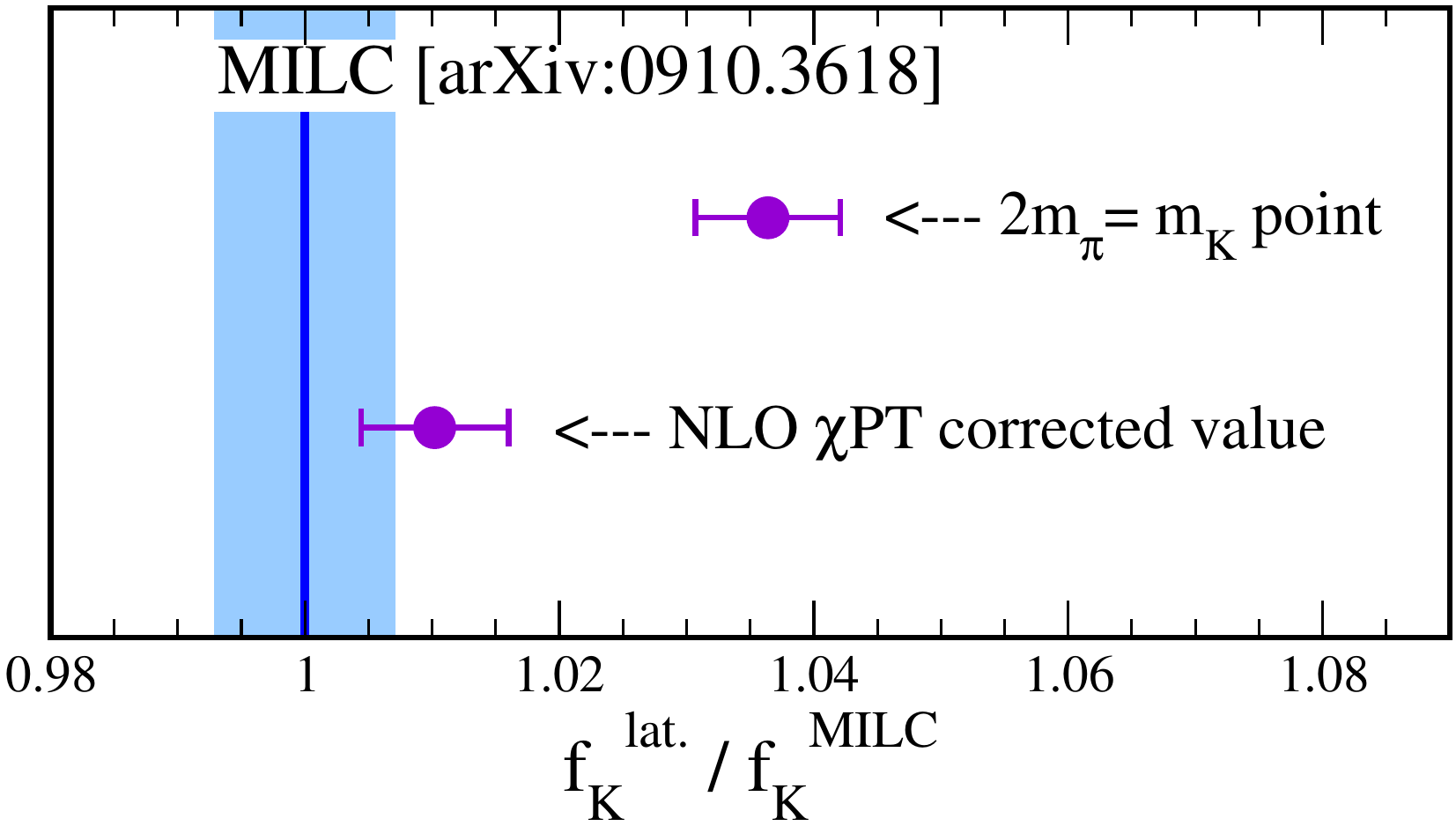}
\vspace{-6mm}
\caption{Demonstration of the method for $f_\pi$ (left plot) and $f_K$ (right plot).  Errors on the circular points only show the statistical errors in the numerator.  Vertical error bands denote the total (statistical plus systematic) uncertainties in $f_\pi$ and $f_K$.  After the interpolation to the unphysical kinematics point $m_K = m_K^{\textrm{phys.}}$ and $m_\pi = m_K^{\textrm{phys.}}/2$, the size of the NLO $\chi$PT correction is below 10\% for $f_\pi$ and below 5\% for $f_K$.}\label{fig:fK_fpi}
\end{center}
\end{figure}

We can test this approach using the known quantities $f_K$ and $f_\pi$; the results are shown in Fig.~\ref{fig:fK_fpi}.  The size of the NLO corrections are quite small (below 10\% for $f_\pi$ and below 5\% for $f_K$), indicating that the systematic uncertainty due to truncating the chiral expansion is under control.

\section{Preliminary determination of Re($A_2$)}

We now use our approach to determine the $(27,1)$ $\Delta I = 3/2$ $K\to\pi\pi$ matrix element, which can be combined with continuum Wilson coefficients~\cite{Ciuchini:1995cd} to obtain Re($A_2$).  

We compute the matrix element in unquenched lattice QCD using asqtad-improved staggered sea quarks and domain-wall valence quarks.  This mixed-action approach shares the primary advantages of both staggered and domain-wall lattice simulations.  We use the publicly-available 2+1 flavor MILC gauge configurations~\cite{Aubin:2004fs}, and simulate with several valence and sea  quark masses.  Although our preliminary analysis only uses two lattice spacings, we have generated data at three lattice spacings ($a \sim 0.06$, 0.09 and 0.12~fm) and will include all of it in a future publication.  Our lightest taste-pseudoscalar sea-sea pion has $m_{\pi,5} = 240$~MeV while our lightest valence-valence pion has $m_\pi = 210$~MeV.  On the $a \sim 0.06$ fm ensembles, the heaviest (taste-singlet) sea-sea pion is also quite light $m_{\pi,I} = 270$~MeV.  This gives us good control over our combined chiral-continuum extrapolation using mixed-action $\chi$PT.  The approximate chiral symmetry of the valence sector ($m_{\textrm{res}} < 3$ MeV on all lattice spacings) makes analysis of the mixed-action simulation data simpler than the purely staggered case. Only two additional parameters appear at 1-loop in the mixed action $\chi$PT expressions for $m_{PS}$, $f_{PS}$, and $B_K$ as compared to the purely domain-wall case~\cite{MAChPT}, and they can both be obtained from spectrum calculations.  Furthermore,  nonperturbative renormalization using the method of Rome-Southampton~\cite{Martinelli:1994ty} can be carried out in a straightforward manner.  Finally, the success of our earlier mixed-action lattice calculation of $B_K$~\cite{Aubin:2009jh} indicates that the mixed-action method is also a good way to determine $K\to\pi\pi$ matrix elements.

\begin{table}
\begin{center}
\caption{Data used for the preliminary determination of Re($A_2$).  The columns show the (i) approximate lattice spacings, (ii) lattice volumes, (iii), nominal up/down ($m_l$) and strange quark ($m_h$) masses in the sea, (iv) corresponding pseudoscalar taste pion mass, (v) partially quenched valence quark masses ($m_x$), (vi) lightest available domain-wall pion mass, and (vii) number of configurations analyzed on each ensemble.}
\bigskip
\label{tab:BK_data}
\footnotesize
\begin{tabular}{llccccr}
\hline\hline
\multicolumn{2}{c}{\qquad} & \multicolumn{2}{c}{sea sector} & \multicolumn{2}{c}{valence sector} \\[-0.5mm]
$a$(fm) & $\left(\frac{L}{a}\right)^3 \times \frac{T}{a}$ & \ $a m_l / am_h$ &  \ $a m_{\pi,5}$ &  \ $am_x$ &  \ $a m_\pi$ & $N_{\rm conf.}$  \\[0.5mm]  \hline

0.06 & $64^3 \times 144$ &  0.0018/0.018 &   0.06678(3) & 0.0026, 0.0108, 0.033 &  0.06376(96) & 96 \\
0.06 & $48^3 \times 144$ &  0.0036/0.018 &   0.09353(7) & 0.0036, 0.0072, 0.0108, 0.033 &  0.07458(76) & 129 \\
\hline


0.12 & $24^3 \times 64$ &  0.005/0.05 &  0.15970(13) & \ 0.007, 0.02, 0.03, 0.05, 0.065 &  0.1718(11) & 218 \\
0.12 & $20^3 \times 64$ &  0.007/0.05 &  0.18887(8) & \ 0.01, 0.02, 0.03, 0.04, 0.05, 0.065 &  0.1968(08) & 279 \\
\hline\hline
\end{tabular}\end{center}\end{table}

\begin{figure}
\begin{center}
\begin{picture}(147,70) 
\put(-1,0){\includegraphics[width=0.48\linewidth]{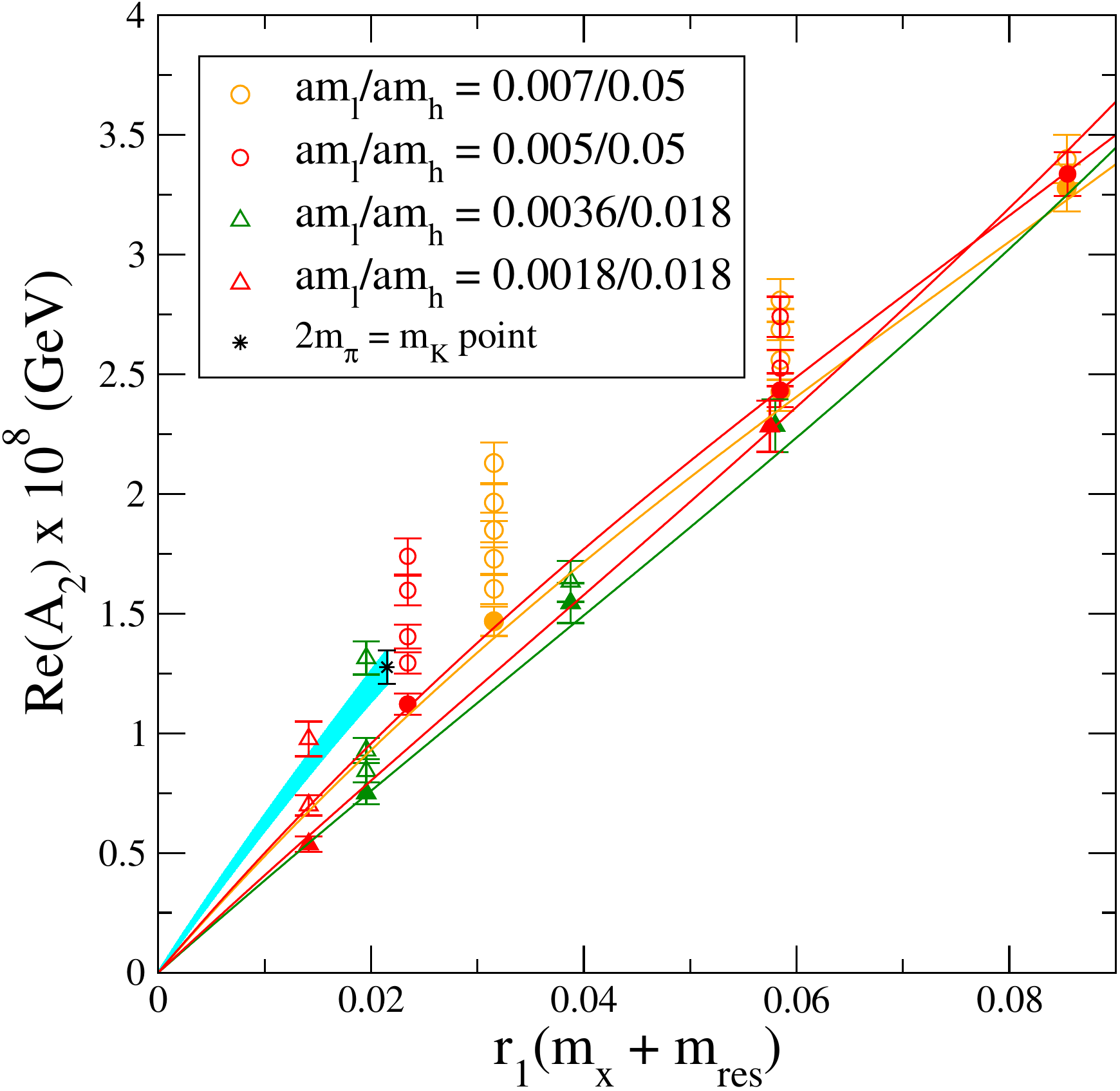}}
\put(75,15){\includegraphics[height=0.23\linewidth]{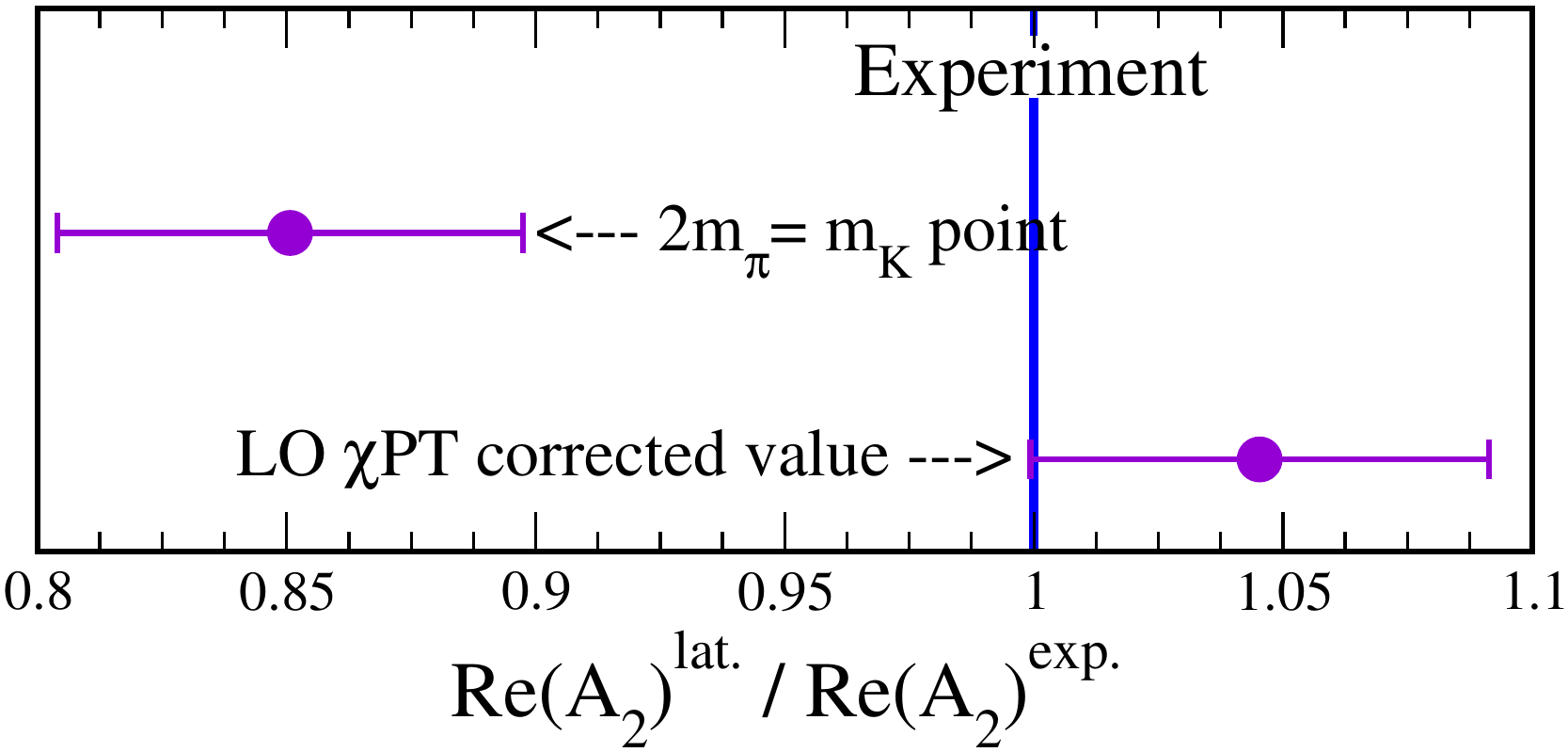}}
\end{picture}
\vspace{-1mm}
\caption{Left plot:  interpolation of Re($A_2$) to the unphysical kinematics point $m_K = m_K^{\textrm{phys.}}$ and $m_\pi = m_K^{\textrm{phys.}}/2$.  Circles (squares) denote $a \sim 0.12$ fm ($a \sim 0.06$ fm) data.  Fit lines correspond to the degenerate mass case and should pass through the filled symbols.  Right plot:  correction of Re($A_2$) to physical kinematics using LO $SU(3)$ $\chi$PT.}
\end{center}\label{fig:ReA2_2pi_interp}
\end{figure}

Figure~\ref{fig:ReA2_2pi_interp} shows the interpolation to the unphysical kinematics point $m_K = m_K^{\textrm{phys.}}$ and $m_\pi = m_K^{\textrm{phys.}}/2$.  Before the interpolation, we adjust the data points by the known 1-loop finite volume corrections, which only depend upon the valence quark masses~\cite{Lin:2001ek}.  The interpolation currently uses LO $\chi$PT supplemented by NLO analytic terms, including a term proportional to $a^2$ so that we can take the continuum limit.  We have finished calculating the mixed-action 1-loop chiral logs, however, and are now working on incorporating them into the fit.  We then correct the unphysical kinematics ``2$\pi$" point to physical kinematics using $SU(3)$ $\chi$PT:
\begin{eqnarray}
	\langle \pi^+ \pi^- | {\mathcal{O}}_i | K^0 \rangle_{\textrm{phys.}} &=& \langle \pi^+ \pi^- | {\mathcal{O}}_i | K^0 \rangle_\textrm{$2\pi$} \times (1 + \delta_\textrm{$\chi$PT}) \,,
\end{eqnarray}
where the correction factor is given by
\begin{eqnarray}
  \delta_\textrm{$\chi$PT} &=& \left( \langle \pi^+ \pi^- | {\mathcal{O}}_{i} | K^0 \rangle_{\textrm{phys.}} -  \langle \pi^+ \pi^- | {\mathcal{O}}_{i} | K^0 \rangle_\textrm{$2\pi$} \right) /  \langle \pi^+ \pi^- | {\mathcal{O}}_{i} | K^0 \rangle_{\textrm{$2\pi$}} \,.
  \end{eqnarray}
Because we have already interpolated to the physical kaon mass, terms in the $\chi$PT expression that are only functions of the kaon mass cancel in the numerator, and the correction is needed only for the short extrapolation from $m_K/2$ to $m_\pi$.  We therefore expect the convergence properties to be better and the truncation errors to be smaller than if we were to extrapolate up to the kaon mass.

At leading order, the expression for the $\Delta I =3/2$ $(27,1)$ $K\to\pi\pi$ matrix element is
 \begin{eqnarray}
	\langle \pi^+ \pi^- | {\mathcal{O}}^{\Delta I = 3/2}_{(27,1)} | K^0 \rangle_{\textrm{LO}} &=& 4 i B_0 f_0 (m^2_K - m^2_\pi) / 3 \,,
\end{eqnarray}
where $f_0$ and $B_0$ are the pion decay constant and $B_K$ in the chiral limits, respectively.  
The $\chi$PT correction factor is then
\begin{eqnarray}
  \delta_\textrm{$\chi$PT}^\textrm{LO} =  \left[ (m_K/2)^2 - m_\pi^2 \right] / \left[ m_K^2 - (m_K/2)^2 \right]
\end{eqnarray}
and is only 23\%, which is much smaller than the chiral correction factors observed by Li and Christ using the standard direct approach~\cite{Li:2008kc}.  Because the low-energy constants $f_0$ and $B_0$ cancel in the ratio, there is no ambiguity regarding the choice of $SU(3)$ LEC's.

The renormalization factor for the $(27,1)$ $\Delta I = 3/2$ operator is the same as $B_K$, so we can use the result for $Z_{B_K}$ from Ref.~\cite{Aubin:2009jh}.  We obtain
\begin{eqnarray}
	\textrm{Re}(A_2) = 1.568(86) \times 10^{-8} ,
\end{eqnarray}	
where the error is statistical only.  This agrees with the experimental measurement, $\textrm{Re}(A_2)_{\textrm{exp}} = 1.50 \times 10^{-8} \textrm{GeV}$~\cite{Buras:1998raa}.  Table~\ref{tab:Error} presents an estimated error budget for $\textrm{Re}(A_2)$.  We assume that the truncation error due to leaving out NLO corrections is half the size of the LO correction, and estimate that total uncertainty in our preliminary result is below 20\%.  This should improve further with the use of our full data set.  The $\chi$PT truncation error and error from the uncertainty in the LEC's will likely also decrease with the use of the NLO $\chi$PT correction factor.  We restrict our lightest valence quark mass to maintain $m_\pi L \gtapprox 3.5$, and estimate that our finite volume errors are a few percent using 1-loop FV$\chi$PT.  We will perform an explicit finite-volume study, however, before publication.   Our preliminary result is renormalized using lattice perturbation theory because we have not yet completed the nonperturbative renormalization on the $a \sim 0.06$~fm ensembles.  We expect these $Z$-factors to be reliable within systematic uncertainties, however, because we find good agreement between $Z_{B_K}$computed nonperturbatively and using lattice perturbation theory on the $a \sim 0.12$ and $a \sim 0.09$ ensembles.

\begin{table}
\begin{center}
\caption{Estimated total error budget for $\textrm{Re}(A_2)$. Each source of uncertainty is given as a percentage.} 
\vspace{-1mm}
\label{tab:total_err}
\begin{tabular}{lr} \\ \hline\hline
uncertainty & $\qquad \textrm{Re}(A_2)$   \\[0.5mm] \hline
statistics & 4.7\%  \\
$\chi$PT truncation error & 12\% \\
uncertainty in leading-order LECs & 4\% \\
discretization errors & 4\% \\
finite volume errors & few percent  \\
renormalization factor & 3.4\%  \\
scale and quark-mass uncertainties & 3\%  \\
Wilson coefficients & few percent \\
\hline
total & less than 20\% \\
\hline\hline
\end{tabular}\label{tab:Error}
\end{center}\end{table}

\end{document}